\documentstyle[aps]{revtex}

\def\be{\begin{equation}}
\def\ee{\end{equation}}
\def\bdm{\begin{displaymath}}
\def\edm{\end{displaymath}}
\def\ba{\begin{array}}
\def\ea{\end{array}}
\def\bea{\begin{eqnarray}}
\def\eea{\end{eqnarray}}

\def\beas{\begin{eqnarray*}}
\def\eeas{\end{eqnarray*}}

\def\d{\partial}

\def\ut#1{\rlap{\lower1ex\hbox{$\sim$}}#1{}}
\def\pb#1{\rlap{\lower1ex\hbox{$\leftarrow$}}#1{}}
\def\real{{\rm I\!R}}

\def\l0{\lambda_0}

\begin{document}
\title{Classical analysis of Bianchi types I and II in Ashtekar variables}
\author{Gabriela Gonzalez \thanks{%
e-mail: gonzalez@suhep.phy.syr.edu }}
\address{Department of Physics, Syracuse University,\\
201 Physics Bldg., Syracuse NY 13244-1130}
\author{Ranjeet S. Tate\thanks{%
e-mail: rstate@phyast.pitt.edu }}
\address{Department of Physics and Astronomy, University of Pittsburgh,\\
100 Allen Hall, 3940 O'Hara St., Pittsburgh PA 15260}
\maketitle

\begin{center}
gr-qc/9412015, UCSBTH-94-44
\end{center}

\begin{abstract}
We solve the complex Einstein equations for Bianchi I and II models
formulated in the Ashtekar variables. We then solve the reality conditions
to obtain a parametrization of the space of Lorentzian solutions in terms of
real canonically conjugate variables. In the Ashtekar variables, the
dynamics of the universe point particle is governed by only a curved
supermetric -- there is no potential term. In the usual metric formulation
the particle bounces off a potential wall in flat superspace. We consider
possible characterizations of this ``bounce'' in the potential-free Ashtekar
variables.
\end{abstract}

\section{Introduction}

The Ashtekar variables for general relativity \cite{aa87} are complex
canonical coordinates on the real phase space, in terms of which Einstein's
equations become low-order polynomials. This has led to several
simplifications and progress on various fronts\cite{newbook,nvreview}.
Classically, the use of the new variables has led to a greater understanding
of the space of selfdual solutions of Einstein's equations \cite{sd}, as
well as of solutions corresponding to degenerate metrics \cite{madv} and
some efforts towards solving the classical constraints \cite{tedj}. In the
canonical quantum theory, the use of new variables has led to the discovery
of large classes of solutions to the quantum scalar constraint equations and
quantum states approximating given 3-geometries. Much of this progress has
resulted from the simplified form of the constraints, in particular, the
scalar constraint is a low-order polynomial, homogeneous and quadratic in the
canonical momenta, albeit the supermetric it defines is in general complex
and non-flat. Thus, dynamical trajectories correspond to null geodesics of
this supermetric and this simplification may allow one to solve the
classical complex equations of motion, or at least to delve deeper into the
structure of the space of solutions.

This progress, however, has come at certain expense. In order to recover
real Lorentzian general relativity, the new variables have to satisfy
certain reality conditions. The configuration variable is a complex $SU(2)$
connection on the spatial 3-manifold $\Sigma $ and the canonically conjugate
momentum is a complex triad (of density weight +1). To recover real
Lorentzian 4-geometries, one requires that the triads be real and that the
real part of the complex $SU(2)$ connection be the spin connection
compatible with the triad.

Can these reality conditions be solved? If one reintroduces the
geometrodynamical variables one obtains the usual formulation which is
manifestly real, but in which the constraints and equations of motion are of
a complicated form. The key question is: Can one first exploit the simple
form of the constraints to solve the complex equations of motion and then
attempt to solve the reality conditions? How does the space of Lorentzian
solutions ``sit'' inside the space of complex solutions? Is there a {\em %
simple} characterization of the Lorentzian solutions in terms of real
canonical variables? Clearly, such a real parametrization of the space of
Lorentzian solutions can be obtained by starting from the geometrodynamical
variables, however, our interest is to learn about the new variables
themselves and the role of the reality conditions, since they appear to have
greater potential for the full theory itself.

This issue -- that of the ``reality'' structure of the space of solutions--
is also likely to be important from the point of view of the quantum theory.
A criterion for the selection of a physical inner product on the space of
solutions to the quantum constraint equation is to require that the reality
conditions on physical observables be represented by Hermiticity conditions
on the corresponding operators \cite{aa:rt}. Functions on the space of
solutions are classical physical observables and thus may provide guides to
constructing quantum Dirac observables.

In the metric variables, the dynamics is that of a ``point particle''
(corresponding to a 3-geometry) moving under the influence of a
potential in a real flat background superspace. In the new variables
the dynamics of the system is governed entirely by the curved
supermetric, and since there is no potential term, the geometric
structure of superspace assumes greater importance, and one would like
to compare the behaviour in terms of connection dynamics to that in
terms of geometrodynamical variables.  Consider the Bianchi models
\cite{aa:jp}, which are certain homogeneous cosmological models of
general relativity, with a finite number of degrees of freedom. In the
geometrodynamical\ variables, the superspace can be taken to be flat
and the particle bounces off potential walls which take different
forms in the different Bianchi models. In Bianchi I the dynamics is
that of a massless relativistic particle. In the simplest nontrivial
case, Bianchi II, each dynamical trajectory bounces once off a single
exponential wall and asymptotically the solution behaves like
different Bianchi I solutions. In Bianchi type IX the potential has a
more complicated shape: it is an expanding triangular well which
confines the particle; and while the system has been shown
analytically to be nonintegrable\cite{lmc}, it is known that the
particle bounces an infinite number of times off these walls \cite
{belinskii}. The dynamics is a sequence of Bianchi I solutions
--during which the particle is largely free of the influence of the
potential-- connected by Bianchi II solutions during which the
particle bounces off one or the other walls. These bounces and their
frequency play an important role in the numerical study of the
dynamics and the presence of chaos in the system\cite{bix}. In terms
of the new variables, even though there is no potential, we expect
that Bianchi II solutions behave asymptotically like different Bianchi
I solutions, mediated by a transition corresponding to the bounce; and
that the Bianchi IX evolution corresponds to an alternating sequence
of Bianchi I and II solutions. Since in Bianchi II there is a single
such transition between different Bianchi I solutions, it serves as a
useful model to look for a characterization of this transition. Such a
``unique'' characterization would be useful for (numerical) studies of
Bianchi IX in the new variables.

In many ways Bianchi types I and II are prototypes for all the solvable
homogeneous cosmologies, and in this paper we will focus our attention on
the more complicated and interesting Bianchi type II. Both the models we
will study in this paper have been solved in the metric variables by other
means. We consider them here to learn something new not so much about the
models themselves but rather about the utility of the new variables.

The plan of the paper is as follows: In section 2 we review the canonical
formulation of the Ashtekar variables in the context of the Bianchi
cosmologies, following Ashtekar and Pullin\cite{aa:jp}, and we discuss some
structure common to all the Bianchi models. Section 3 is devoted to a study
of the Bianchi type I model. While this model is quite trivial and has been
quite thoroughly solved and understood in the new variables, the motivation
here is to solve Bianchi I via a process which will be useful to solve
Bianchi II, and to understand and familiarize ourselves with that process.
In the first part of section 4 we will find the general solution to the
classical complex Hamiltonian equations of motion for Bianchi II, and find
that the space of solutions is parametrized by two canonically conjugate
pairs of variables. In the second part of section 4 we will require that the
triads be real throughout the evolution for all solutions. While the reality
conditions are complicated in terms of the triad and the connection, they
lead to simple conditions on the chosen reduced phase space\ coordinates:
all the parameters turn out to be real, and thus the resulting reduced phase
space description is simple. In section 5 we characterize the ``bounce''
during the transition from one asymptotic Bianchi I solution to another in
terms of certain functions of the connection components.

\section{Preliminaries}

Let us recall the Ashtekar variables for general relativity \cite
{newbook,nvreview}. These are the canonically conjugate pair $(\widetilde{E}%
_i^a,A_a^i)$ of complex coordinates on the phase space of real general
relativity, where $\widetilde{E}{}_i^a$ is the densitized triad and $A_a^i$
is a ${\rm C}\llap{\vrule height7.1pt width1pt depth-.4pt\phantom t} SU(2)$
connection. These variables are related to the usual canonical metric
variables via

\begin{equation}
\widetilde{E}{}_i^a\widetilde{E}{}^{bi}=\widetilde{\widetilde{q}}{}^{ab}\;%
{\rm and}\;A_a^i=\Gamma _a^i-iK_a^i,  \label{avtogmd}
\end{equation}
where $\widetilde{\widetilde{q}}{}^{ab}$ is the density weight $+2$
contravariant 3-metric on the spatial slices, $\Gamma _a^i$ is the spin
connection of the triad and $K_a^i$ is the (triad component of the)
extrinsic curvature of the spatial manifold.

Now let us consider the spatially homogeneous Bianchi models. The reduction
of the Ashtekar variables to these cosmologies has been carried out
previously\cite{aa:jp,kodama}. Here we will review the canonical
formulation, following the conventions and notation of Ashtekar and Pullin%
\cite{aa:jp}. The Bianchi models are spatially homogeneous models which
admit a three dimensional isometry group which acts simply and transitively
on the preferred homogeneous spatial slices. One introduces a basis (and its
dual) of group invariant 1-forms $\omega ^I, I=1-3$ which satisfy $d\omega
^I=- {\textstyle{\frac{1}{2}}} C^I{}_{JK}\omega ^J\wedge \omega ^K$, where $%
C^I{}_{JK}$ are the structure constants of the Lie algebra associated with
the Bianchi group. Components of homogeneous tensors in this basis are
spatial constants; e.g.\ the 4-metric in this basis is
\[
ds^2=-N^2(t)dt^2+\sum_{IJ}q_{IJ}(t)\omega ^I\bigotimes \omega ^J,
\]
where the lapse function $N(t)$ and the 3-metric components $q_{IJ}$ are
constant on the homogeneous spatial slices and are thus functions only of
the time $t$.

Class A models ---distinguished by the vanishing of the trace of $C$---
readily admit a Hamiltonian formulation \cite{aa:sam}. The structure
constants for the Class A models can be written as
\begin{equation}
C^I{}_{JK}=n^{(I)}\epsilon _{IJK}\;(\mbox{no sum over }\quad I),
\label{struct}
\end{equation}
where $\epsilon _{IJK}$ is the totally antisymmetric tensor and the set of
numbers $\{n^{(I)}\}$ ---each of which can take the values $0,\pm 1$--- are
used to classify the different models. In this paper we are interested in
the Bianchi types I and II, which are both in Class A.

For simplicity we will further confine our attention to the diagonal gauge, in
which the off-diagonal components of the spatially homogeneous triad and
connection vanish. The resulting phase space is 6 dimensional. In this
diagonal gauge, the 6 canonically conjugate phase space variables are $%
\{A_I,-iE_I\}$, with $I=1,2,3$ and the symplectic form is given by
\begin{equation}  \label{genss}
\Omega =-i\sum_IdE_I\wedge dA_I.
\end{equation}
For future reference note that the relationship between the 3-metric and the
triads for the Bianchi cosmologies is
\begin{equation}  \label{cantrans}
q_I=\left| \frac{E_1E_2E_3}{E_I^2}\right| .
\end{equation}

In the diagonal gauge the diffeomorphism and Gauss constraints vanish
identically, and we are left with the scalar constraint:
\begin{equation}  \label{gensc}
\begin{array}{rcl}
{\cal S} & \equiv & G_N\,G^{IJ}(A,n)\,p_Ip_J \\
& = & -G_N\left[ (A_IA_J-\delta _{IJ}(A_I)^2)+\frac 1{G_N}n^{(K)}(\epsilon
_{KIJ})^2A_K\right] (-iE_I)(-iE_J),
\end{array}
\end{equation}
where $G_N$ is Newton's constant, $G^{IJ}$ is the (contravariant)
supermetric on the configuration space of connections, and varies from one
Bianchi model to another through its dependence on $n^{(I)}$. Note that we
have made a choice for the real lapse function: $\rlap{\lower1ex\hbox{$%
\sim$}}N{}=1$, thus $N=(E_1E_2E_3)^{1/2}$.

In general relativity, classical dynamics is generated by the (vanishing)
scalar constraint. Let $\lambda $ be an affine parameter along the orbits
generated by the scalar constraint. Then the evolution of any function on
phase space\ is given by $\dot f:=df/d\lambda =\{f,{\cal S}\}$. Since the
scalar constraint function in the new variables consists of only a kinetic
term quadratic in the momenta, and is constrained to vanish, the dynamics it
generates is that of a massless free particle in a curved spacetime with a
(super)metric given by (\ref{gensc}). Thus, in configuration space the
dynamical trajectories are simply the null geodesics of the above
supermetric. Even though the supermetric is complex, its geodesics and its
null directions are both well defined.

As we can see from the canonical transformation (\ref{avtogmd}), the
coordinates $(E_I,A_I)$ are in general complex. In order to recover the real
Lorentzian solutions, we have to impose ``reality conditions'' on the
canonical variables. These conditions are:

\begin{equation}  \label{genrc}
\begin{array}{rl}
E_I & \in {\rm I\!R} \\
Re(A_I) & =-{\displaystyle \frac{1}{2G_N}}{\displaystyle \frac{\d}{
\d E_I}}\left( \sum_J{\displaystyle \frac{E_1E_2E_3n^{(J)}}{E_J^2}}
\right),
\end{array}
\end{equation}
where the $n^{(I)}$ defined in (\ref{struct}) distinguish the various Class
A Bianchi models. Both the scalar constraint (\ref{gensc}) and the
symplectic form (\ref{genss}) are real when evaluated on triads and
connections satisfying the above reality conditions. Since the triads are
also required to be real, the parameter along the real evolution generated
by the (real) scalar constraint should also be real. We see that the reality
conditions in the form $E_I\in {\rm I\!R},\dot E_I\in {\rm I\!R}$ --- where
the dot $^{.} $ indicates derivative with respect to a real parameter along
the orbits generated by a scalar constraint smeared with a real lapse
$\rlap{\lower1ex\hbox{$\sim$}}N{}=1$ -- are completely equivalent to the
reality conditions in the ``canonical'' form (\ref{genrc}).

In the rest of this paper we will consider the Bianchi type I and II models,
which are specified by $n^{(I)}=0$ and $n^{(1)}=1,n^{(2)}=n^{(3)}=0$
respectively.

\section{Bianchi I}

Now let us consider Bianchi type I, where the $n^{(I)}=0$. The supermetric
is then
\begin{equation}  \label{I:smetric}
G^{IJ}=-{\displaystyle \frac{1}{2}}\left[
\begin{array}{ccc}
0 & A_1A_2 & A_1A_3 \\
A_1A_2 & 0 & A_2A_3 \\
A_1A_3 & A_2A_3 & 0
\end{array}
\right] .
\end{equation}

In order to simplify the calculations we will change coordinates on the
space of connections, $\{A_I\}\rightarrow \{x^i=t,x,y\}$ as follows:
\begin{equation}  \label{I:txy}
\begin{array}{ccc}
t & = & A_2A_3 \\
x & = & G_NA_2A_3/A_1 \\
y & = & A_2/A_3.
\end{array}
\end{equation}
In these coordinates the metric $G$ is now diagonal:
\begin{equation}  \label{I:diagmetric}
G^{ij}=diag(-t^2,x^2,y^2),
\end{equation}
and the connection components are
\begin{equation}  \label{I:conn}
\begin{array}{ccc}
A_1 & = & G_Nt/x \\
A_2 & = & \sqrt{ty} \\
A_3 & = & \sqrt{t/y}.
\end{array}
\end{equation}
Note that there are many ways to diagonalize the supermetric, this
particular choice of coordinates has been made in order to follow as closely
as possible the change of coordinates which simplifies the calculations for
the Bianchi II model.

Since we are interested in the evolution of the triads, and will impose
reality conditions\ on the triads, let us express them in terms of the new
coordinates and momenta. Let $p_i=\{p_t,p_x,p_y\}$ be the momenta
canonically conjugate to the new coordinates $t,x,y$ respectively, so that
the triads are given by
\begin{equation}  \label{I:triads}
\begin{array}{rcl}
E_1 & = & -\frac i{A_1}xp_x \\
E_2 & = & \hphantom{-}\frac i{A_2}(tp_t+xp_x+yp_y) \\
E_3 & = & \hphantom{-} \frac i{A_3}(tp_t+xp_x-yp_y),
\end{array}
\end{equation}
and the symplectic form is
\begin{equation}  \label{ssnew}
\Omega=\sum_i dp_i\wedge dx^i.
\end{equation}

{}From (\ref{I:txy}) we see that under the transformation $(A_2,A_3)
\rightarrow (-A_2,-A_3)$ the new coordinates are left unchanged, and there
is the related ambiguity in the square roots in (\ref{I:conn}). We will
resolve this ambiguity by taking the principal value of the square roots,
and thus work with a quarter of the original phase space. The other parts
are recovered by choosing the other signs of the square roots. We are
justified in considering these ``quadrants'' separately since they are
dynamically disconnected, i.e.\ the Hamiltonian vector field of the
constraint is tangential to the boundary between any two of them.
Furthermore, the spacetime geometries that result from the different
choices of sign are in fact the same.

Let us solve the (complex) equations of motion. For Bianchi type I the null
geodesics are easy to find since the metric has three Killing vector fields:
$K_i=x^i\partial /\partial x^i$. Let $\lambda \in{\rm C}%
\llap{\vrule
height7.1pt width1pt depth-.4pt\phantom t}$ be a parameter along the orbits
generated by the scalar constraint, and let $x^i(\lambda )$ be an orbit in
configuration space. Then, associated with the Killing vectors we have three
conserved quantities :
\begin{equation}  \label{ciI}
\begin{array}{rcccl}
c_i & = & x^ip_i & = & G_{jk}\;K_i^j\;\dot x{}^k \\
\mbox{ i.e. }\quad \{c_t,c_x,c_y\} & = & \{tp_t,xp_x,yp_y\} & =
& \{-\dot t/t, \dot x/x,\dot y/y\},
\end{array}
\end{equation}
We require the trajectories to be null, in order to satisfy the scalar
constraint. This yields a condition on the above constants:

\begin{equation}  \label{cteqn}
-c_t^2+c_x^2+c_y^2=0.
\end{equation}

We can directly solve the equations of motion (\ref{ciI}) for the orbits to
obtain

\begin{equation}  \label{I:orbits}
\begin{array}{rcl}
t(\lambda ) & = & t_0e^{-c_t\lambda } \\
x(\lambda ) & = & e^{c_x(\lambda -\lambda _0)} \\
y(\lambda ) & = & y_0e^{c_y\lambda },
\end{array}
\end{equation}
where the complex constants $(t_0,\lambda _0,y_0)$ correspond to the initial
values of the coordinates $(t,x,y)$ respectively (We have chosen $\lambda _0$
as a parameter instead of $x_0$ following the most convenient choice for
Bianchi II).

If we substitute (\ref{I:orbits}) into (\ref{I:conn}) and (\ref{I:triads}),
we obtain the general complex solutions to the scalar constraint:\ we still
have to impose the reality conditions. But before we proceed to the real
solutions, we will first calculate the symplectic form on the complex
reduced phase space $\hat \Gamma $. We will do this in two steps. First, we
evaluate $\Omega $ on the complex solutions (effectively taking as phase
space coordinates the orbit parameters and the affine parameter along the
orbits $\{c_t,c_x,c_y,t_0,y_0,\lambda \}$):

\begin{equation}
\begin{array}{rcl}
\Omega |_{x_i(\lambda )} & = & \sum_idc_i\wedge d\ln x^i \\
& = & dc_t\wedge d\ln t_0+dc_y\wedge d\ln y_0+d\lambda \wedge
(c_tdc_t-c_ydc_y-c_xdc_x).
\end{array}
\end{equation}
Next we pull-back to the constraint surface given by (\ref{cteqn}), to
obtain the symplectic form on the reduced phase space
\begin{equation}  \label{I:redsymp}
\hat \Omega =dc_t\wedge d\ln t_0+dc_y\wedge d\ln y_0.
\end{equation}
Therefore we see that $\{c_t,c_y,\ln t_0,\ln y_0\}$ is a set of (complex)
canonical coordinates on the complex reduced phase space. We solve (\ref
{cteqn}) for the parameter $c_x$ as a function of the free parameters $%
c_t,c_y$: $c_x=pr.val.\sqrt{c_t^2-c_y^2}$. We will introduce in the
solutions the discrete parameter $\epsilon =\pm 1$ to capture the ambiguity
in sign of the solution to (\ref{cteqn}).

We now impose the reality conditions\ (\ref{genrc}) on the complex solutions
and find the phase space of the real degrees of freedom. Recall from the
discussion in section 2 that this is equivalent to the requirement that the
triads $E_I$ be real for all ``real'' evolution, and that the parameter
along the real trajectories should be real. Hence it is the real part of $%
\lambda $ which is the parameter along the real dynamical trajectories: if
we write $\lambda =\tau +i\phi $, then $\tau $ is the time, and $\phi $ is a
fixed real orbit parameter. The remaining reality conditions will lead in
turn to conditions on the orbit parameters $\{c_t,c_y,\ln t_0,\ln y_0,\phi
\} $. Let $\lambda _0=\tau _0+i\phi _0$, and identify $\tau _0$ as the real
initial time.

We define new orbit parameters
\begin{equation}  \label{I:newpar}
T_0=t_0\exp (-ic_t\phi )\quad \mbox{ and }\quad Y_0=y_0\exp (ic_y\phi ).
\end{equation}
{}From the expression (\ref{I:triads}) and the solutions (\ref{I:orbits}) we
obtain the solutions for the triads
\begin{equation}  \label{I:triadsol}
\begin{array}{rll}
E_1 & =-i{\displaystyle \frac{\epsilon c_x}{A_1}} & =-i {\displaystyle \frac{%
\epsilon c_x}{G_NT_0}}\;\exp ((c_t+\epsilon c_x)\tau -\epsilon
c_x\tau_0+i\epsilon c_x(\phi-\phi_0) ) \\
E_2 & =i{\displaystyle \frac{c_t+\epsilon c_x+c_y }{A_2}} & =i {%
\displaystyle \frac{c_t+\epsilon c_x+c_y}{\sqrt{T_0Y_0}}}\exp ((c_t-c_y)\tau
/2) \\
E_3 & =i{\displaystyle \frac{c_t+\epsilon c_x-c_y}{A_3}} & =i(c_t+\epsilon
c_x-c_y)\sqrt{\frac{ Y_0 }{T_0}}\exp ((c_t+c_y)\tau /2).
\end{array}
\end{equation}

Let us impose the rest of the reality conditions. Since we only have
exponential functions of $\tau $, we see that $d\ln E_I/d\tau $ are
real if and only if all $c_i$ are real (and hence $c_t^2\ge
c_y^2$). Next, we see that $E_2$ and $E_3$ are real if and only if in
addition $T_0,Y_0$ are real and satisfy $T_0Y_0<0$. Then, $E_1$ will
be real only if $\exp (i\epsilon c_x(\phi -\phi _0))=\pm i$, from
which we conclude that $\cos (c_x(\phi -\phi _0))=0$.

Collecting all reality conditions, we conclude that the triads are real if
and only if
\begin{equation}  \label{I:realconds}
\begin{array}{c}
c_t,c_y\in {\rm I\!R}\quad {\rm and }\quad c_t^2\ge c_y^2 \\
T_0,Y_0\in {\rm I\!R}\quad {\rm and }\quad T_0Y_0<0 \\
{\rm and }\qquad \cos (c_x(\phi -\phi _0))=0 .
\end{array}
\end{equation}
Note that $\tau _0$ corresponds to a choice of the initial value of $x$ and
has been left as a free parameter, but $\phi $ and $\phi _0$ have
disappeared as orbit parameters, fixed by the reality conditions. The real
solution for the triads is then simply

\begin{equation}  \label{reale1}
\begin{array}{l}
E_1=\pm {\displaystyle \frac{\epsilon c_x}{G_NT_0}}\;\exp ((\epsilon
c_x+c_t)\tau -\epsilon c_x\tau _0) \\
E_2= {\displaystyle \frac{c_t+\epsilon c_x+c_y}{\sqrt{-T_0Y_0}}}\exp
((c_t-c_y)\tau /2) \\
E_3=(c_t+\epsilon c_x-c_y)\sqrt{-\frac{Y_0}{T_0}}\exp ((c_t+c_y)\tau /2).
\end{array}
\end{equation}
Note that the sign ambiguities in the above solution for the triads lead to
distinct triad solutions. However, the 4-dimensional spacetime geometries
(quadratic in the triads) are unaffected by the choice of sign and thus the
solutions are physically equivalent.

When the reality conditions are satisfied, the connections are pure
imaginary:
\begin{equation}  \label{I:realconn}
\begin{array}{ccc}
A_1 & = & \pm iG_NT_0e^{\epsilon c_x\tau _0}\exp (-(c_t+\epsilon c_x)\tau )
\\
A_2 & = & i \sqrt{-T_0Y_0}\exp \left( -(c_t-c_y)\tau /2\right) \\
A_3 & = & i\sqrt{ -T_0/Y_0}\exp \left( -(c_t+c_y)\tau /2\right).
\end{array}
\end{equation}
All the connections have a simple exponential behavior, vanishing (at
different rates) as $c_t\tau \rightarrow \infty $. Note that there are sign
ambiguities in all three connection components arising due to either the
phase $\epsilon c_x(\phi -\phi _0)$ or the square root. These sign
ambiguities also lead to different solutions which however correspond to the
same spacetime geometry.

{}Finally, the pull-back of the symplectic form to the space of real
solutions still looks very much as in (\ref{I:redsymp}):
\begin{equation}  \label{realsymp}
\hat \Omega =dc_t\wedge d\ln T_0+dc_y\wedge d\ln Y_0.
\end{equation}
Thus, the chosen coordinates on the reduced phase space are real as well as
canonically conjugate.

{}From the above solutions and (\ref{cantrans}), one can obtain the
components of the metric. The metric is the Kasner solution\cite{kasner}:
\begin{equation}  \label{bImet}
ds^2=-dt^2+t^{2p_1}\;dx^2+t^{2p_2}\;dy^2+t^{2p_3}\;dz^2 ,
\end{equation}
where we have rescaled the time function
\begin{equation}  \label{kasnertime}
e^{(\epsilon c_x+2c_t)\tau }\rightarrow t
\end{equation}
and the set of Kasner parameters ${\bf p=\{}p_1,p_2,p_3\}$ are functions of
the orbit parameters
\begin{equation}  \label{kasnerpmts}
\{c_t,c_x,c_y\}\rightarrow {\bf p=}\{-\epsilon c_x,\epsilon
c_x+c_y-c_t,\epsilon c_x-c_y-c_t\}/(\epsilon c_x+2c_t).
\end{equation}
The properties of the Kasner parameters $\sum p_i^2=\sum p_i=1$ are derived
from (\ref{cteqn}). Belinskii {\it et al}\cite{belinskii} introduced a
single parameter $u\in [0,1)$ as a solution to the above constraints, where $%
(1+u+u^2){\bf p}(u)=\{-u,1+u,u(1+u)\}$. In terms of the parameters we have
used, $u^2=(|c_t|-|c_y|)/(|c_t|+|c_y|)$ for the ratio $\left| c_y/c_t\right|
$ in the range $[0,1]$. The solutions with parameters $\left| c_t/c_y\right|
=1$ correspond to flat Rindler spacetime\ (i.e., Minkowski spacetime\ in an
accelerated frame). The ``trivial'' solution with all $c_i=0$ is, of course,
Minkowski spacetime. In the non-trivial cases, the initial singularity is
approached in the limit $-c_t\tau \rightarrow \infty $.

\section{Bianchi II}

In this section we will solve the classical dynamics of the Bianchi type II
model, which is specified by the structure constants $n^{(1)}=1=C^1{}_{23}$.
We will repeat the procedure step-by-step as for Bianchi type I.

The Bianchi type II supermetric is

\begin{equation}  \label{II:smetric}
G^{IJ}=-{\displaystyle \frac{1}{2}}\left[
\begin{array}{ccc}
0 & A_1A_2 & A_1A_3 \\
A_1A_2 & 0 & A_2A_3 + A_1/G_N \\
A_1A_3 & A_2A_3+A_1/G_N & 0
\end{array}
\right].
\end{equation}

As for Bianchi type I, we will change coordinates to a set adapted to the
Killing symmetries of the supermetric, $\{A_I\}\rightarrow\{x^i=t,x,y\}$ as
follows:

\begin{equation}  \label{II:txy}
\begin{array}{rcl}
t & = & A_2A_3-A_1/G_N \\
x & = & G_NA_2A_3/A_1 \\
y & = & A_2/A_3.
\end{array}
\end{equation}
In these coordinates the metric $G$ is now diagonal:
\begin{equation}  \label{II:diagmetric}
G^{ij}=diag(-\frac{x}{x-1}\cdot t^2,x(x-1), \frac{x+1}{ x }\cdot y^2).
\end{equation}

The connection components are expressed in terms of the Killing coordinates
by
\begin{equation}  \label{II:conn}
\begin{array}{rcl}
A_1 & = & G_Nt/(x-1) \\
A_2 & = & \sqrt{txy/(x-1)} \\
A_3 & = & \sqrt{tx/(y(x-1))}.
\end{array}
\end{equation}
(The ambiguity in the sign of $A_2,A_3$ we resolve as for Bianchi type I;
see the discussion after (\ref{I:conn}).) Let $p_i=\{p_t,p_x,p_y\}$ be the
momenta canonically conjugate to the new coordinates $t,x,y$ respectively.
Then the triads are given by
\begin{equation}  \label{II:triads}
\begin{array}{rcl}
E_1 & = & -{\displaystyle \frac{i}{A_1}}( {\displaystyle \frac{tp_t}{x-1}}
+xp_x) \\
E_2 & = & {\displaystyle \frac{i}{A_2}}({\displaystyle \frac{x}{x-1}}
tp_t+xp_x+yp_y) \\
E_3 & = & {\displaystyle \frac{i}{A_3}}({\displaystyle \frac{x}{x-1}}
tp_t+xp_x-yp_y).
\end{array}
\end{equation}

{}For Bianchi type II the supermetric has only two linearly independent
Killing vector fields. However, since the space is 3-dimensional, the system
is separable, and we will be left with three ordinary differential equations
to solve. The two Killing vector fields are: $K_t=t\partial /\partial t$ and
$K_y=y\partial/\partial y$. As before, let $\lambda \in{\rm C}%
\llap{\vrule
height7.1pt width1pt depth-.4pt\phantom t}$ be a parameter along the orbits
generated by the scalar constraint, and let $x^i(\lambda )$ be an orbit in
configuration space. Then, associated with the Killing vectors we have two
conserved quantities :
\begin{equation}  \label{II:ci}
\begin{array}{ccc}
c_t & =tp_t & =tG_{tt}\;\dot t= {\displaystyle \frac{1-x}{x}}\;{\displaystyle
\frac{\dot t}{t}} \\
c_y & =yp_y & =yG_{yy}\;\dot y={\displaystyle \frac{x}{x+1}}\;{\displaystyle
\frac{\dot y}{y}}.
\end{array}
\end{equation}

Let us define $c_x$ similar to the Bianchi I parameter:
\begin{equation}
c_x^{}=pr.val.\sqrt{c_t^2-c_y^2}.
\end{equation}
The constraint (i.e.\ that the trajectories be null) yields an equation for $%
\dot x$
\begin{equation}  \label{cxeqn}
\dot x^2=x^2c_x^2+c_y^2.
\end{equation}
This equation can be readily solved for $x(\lambda )$:
\begin{equation}  \label{II:xsol}
x(\lambda )=\epsilon \frac{c_y}{c_x}\sinh\left(c_x(\lambda-\l0)\right),
\end{equation}
where $\epsilon =\pm 1$ corresponds to the two possible choices of sign in
the square root of (\ref{cxeqn}).

Substituting the above solution $x(\lambda )$ in the equations (\ref{II:ci})
for the conserved quantities we obtain ordinary linear differential
equations for $t(\lambda )$ and $y(\lambda )$ which can be solved to yield
the complete equations of motion:
\begin{equation}  \label{II:orbits}
\begin{array}{rl}
t(\lambda )= & t_0e^{-c_t\lambda } {\displaystyle \frac{c_x}{c_y}}{%
\displaystyle \frac{\dot x+xc_t}{\dot x-c_t}} \\
&  \\
= & t_0e^{-c_t\lambda } {\displaystyle \frac{c_x\cosh\left(c_x(\lambda-\l0)%
\right)+c_t\sinh\left(c_x(\lambda-\l0)\right)}{c_y\cosh\left(c_x(\lambda-%
\l0)\right)-\epsilon c_t}} \\
&  \\
y(\lambda )= & y_0e^{c_y\lambda } {\displaystyle \frac{\dot x-c_y}{xc_x}} =
y_0e^{c_y\lambda }{\displaystyle \frac{\cosh\left(c_x(\lambda-\l0)\right)-%
\epsilon }{\sinh\left(c_x(\lambda-\l0)\right)}},
\end{array}
\end{equation}
where the complex constants $t_0,y_0$ correspond to the initial values of
the coordinates $t,y$ respectively.

A lengthy calculation shows that the symplectic structure evaluated on the
solutions is simply
\begin{equation}  \label{II:ss}
\hat \Omega = dc_t\wedge \ln t_0+dc_y\wedge \ln y_0,
\end{equation}
from which it is clear that $\{c_t,c_y,\ln t_0,\ln y_0\}$ is a set of
canonical coordinates on the complex reduced phase space. The calculation of
the symplectic form on the reduced phase space is considerably simplified
when we note that the pull-back of the symplectic structure to the
constraint surface has two properties:{\it i)} it is degenerate --with the
degenerate direction along the Hamiltonian vector field of the constraint--
and {\it ii)} its Lie derivative along the above vector field vanishes.
Hence, there can be no terms in $\hat \Omega $ containing $d\lambda $, and
the coefficients of the remaining terms will be independent of $\lambda $.
Thus, we can simply evaluate the symplectic structure on some $\lambda
=const $ cross-section of the constraint surface.

We now impose the reality conditions\ (\ref{genrc}) on the complex solutions
and find the reduced phase space of the real degrees of freedom. As before
we can identify $\tau=Re(\lambda)$ with the real time, and require that the
triads and their derivatives with\ respect\ to\ $\tau$ should be real. It is
more convenient to (partially) solve the reality conditions in terms of $
E_2\cdot E_3$ and $E_2/E_3$. First, we substitute for the momenta $p_i$ in
terms of the velocities $\dot{x}{}^i$ in the expressions (\ref{II:triads}).
Various simplifications then lead to the following expressions:
\begin{equation}  \label{II:e2e3}
\begin{array}{ccc}
E_r(\tau ) & := & E_2/E_3={\displaystyle \frac{1}{Y_0}} {\displaystyle \frac{
c_x}{c_t-c_y}}e^{-c_y\tau } \\
E_p(\tau ) & := & E_2\cdot E_3=-{\displaystyle \frac{2}{T_0}}c_xc_ye^{c_t\tau
},
\end{array}
\end{equation}
where we have defined the new orbit parameters $T_0,Y_0$ exactly as for
Bianchi I (\ref{I:newpar}). Now $d\ln (E_r))/d\tau $ and $d\ln (E_p))/d\tau $
are real if and only if
\begin{equation}
c_y,c_t\in {\rm I\!R},
\end{equation}
and the reality of $E_p,E_r$ is then equivalent to
\begin{equation}
c_x/Y_0,c_x/T_0\in {\rm I\!R},
\end{equation}
where $c_x$ is either real or imaginary. In order to guarantee that $E_2,E_3$
are themselves real, and not possibly pure imaginary, we also require that $
E_r\cdot E_p\ge 0$, this condition on the orbit parameters is

\begin{equation}  \label{II:quart}
c_y(c_t+c_y)Y_0 T_0\le0.
\end{equation}

Now let us consider the reality of $E_1$. With the above reality conditions\
on $E_2,E_3$ imposed, the solution for $E_1$ is
\begin{equation}  \label{II:e1}
E_1(\tau )=\frac{-i}{T_0}e^{c_t\tau }\left( c_t\sinh \left( c_x(\tau -\tau
_0+i(\phi -\phi _0)\right) -c_x\cosh \left( c_x(\tau -\tau _0+i(\phi -\phi
_0)\right) \right) ,
\end{equation}
where, as for Bianchi type I we have defined $\lambda _0=\tau _0+i\phi _0$.
We will impose the reality conditions on $E_1,\dot E_1$ at $\tau =\tau _0$.
Using the previously solved reality conditions on the orbit parameters, the
remaining reality conditions are reduced to
\begin{equation}  \label{II:E1real}
\begin{array}{rcl}
E_1(\tau _0)\in {\rm I\!R} & \Longleftrightarrow & {\displaystyle \frac{c_t}{%
c_x}}\sin (c_x(\phi -\phi _0))+i\cos (c_x(\phi -\phi _0))\in {\rm I\!R} \\
\mbox{ and }\quad \dot E_1(\tau _0)\in \quad{\rm I\! R} &
\Longleftrightarrow & c_x\sin
(c_x(\phi -\phi _0))+ic_t\cos (c_x(\phi -\phi _0))\in\quad {\rm I\! R}.
\end{array}
\end{equation}
Now we know that $c_x$ is either real or pure imaginary; in either case the
first terms in the above equations are real, and the second terms are pure
imaginary and hence should vanish. Thus, we should have $\cos (c_x(\phi
-\phi _0))=0$, which has {\em no} solutions for imaginary $c_x$, and we
conclude that $c_x\in {\rm I\!R}$ and thus $c_t^2\ge c_y^2$. Putting this back
into \ref{II:E1real}, we see that these (necessary) conditions are
sufficient to guarantee that $E_1$ is real for all $\tau $.

Collecting all the reality conditions, we conclude that the solutions for
the triads in Bianchi type II will be real if and only if
\begin{equation}  \label{II:realconds}
\begin{array}{c}
c_t,c_y\in {\rm I\!R}\quad {\rm and }\quad c_t^2\ge c_y^2 \\
T_0,Y_0\in {\rm I\!R}\quad {\rm and }\quad c_tc_yT_0Y_0\le 0 \\
{\rm and } \quad \cos (c_x(\phi -\phi _0)))=0.
\end{array}
\end{equation}
(In the inequality (\ref{II:quart}) the sign of $(c_t+c_y)$ is determined by
the sign of $c_t$. Note also that $\tau _0$, which corresponds to the
initial value of $x$, is left undetermined by any of the reality conditions,
and is a free parameter.)

The triads evaluated on the real trajectories are:
\begin{equation}  \label{II:realtriads}
\begin{array}{rcl}
E_1 & = & \pm \frac{c_y}{G_NT_0}\exp (c_t\tau )\cosh (c_x(\tau -\tau
_0)-\phi ) \\
&  &  \\
E_2 & = & \sqrt{\frac{-2}{T_0Y_0}c_y(c_t+c_y)}\exp (c_t-c_y)\tau /2 \\
&  &  \\
E_3 & = & \sqrt{\frac{-2Y_0}{T_0}c_y(c_t-c_y)}\exp (c_t+c_y)\tau /2,
\end{array}
\end{equation}
where $\phi =$arctanh$(c_x/c_t)$.

If we evaluate the metric, we find the solution quoted by Belinskii {\it et
al}\cite{belinskii} (as an approximate solution for a ``Kasner epoch''
transition in Bianchi IX) with the same identification of parameters we made
in Bianchi I in (\ref{kasnerpmts}), with $\epsilon =+1$. It is also, of
course, the solution first quoted for Bianchi II models by Taub\cite{taub},
with a similar identification of parameters $k=c_x,c_1=c_t-c_y$ and $%
c_2=c_t+c_y$.

The connections $A_I$ evaluated on the real trajectories are :

\begin{equation}  \label{II:realconn}
\begin{array}{rcl}
A_1 & = & -G_N(c_x/c_y)T_0\exp (-c_t\tau )\left( 1+i\epsilon f(\tau )\right)
^{-1} \\
&  &  \\
A_2^2 & = & T_0Y_0\exp (-(c_t-c_y)\tau )(1-i\epsilon \sinh (c_x(\tau -\tau
_0)))\left( 1-i\epsilon f(\tau )\right) ^{-1} \\
&  &  \\
A_3^2 & = & -(T_0/Y_0)\,\exp (-(c_t+c_y)\tau )(1+i\epsilon \sinh (c_x(\tau
-\tau _0)))\left( 1-i\epsilon f(\tau )\right) ^{-1}.
\end{array}
\end{equation}
where $f(\tau )=\delta \sinh (c_x(\tau -\tau _0)-\phi )$, with $\delta
=sgn(c_tc_y)$ .

As $\tau \rightarrow \pm \infty $, the connections behave respectively
as:
\begin{equation}  \label{limconn}
\begin{array}{rcl}
A_1 & \rightarrow & -i\epsilon G_NT_0\exp (\pm c_x\tau _0)/(1\mp
c_t/c_x)\,\,\exp (-c_t\pm c_x\tau ) \\
&  &  \\
A_2 & \rightarrow & i \sqrt{\left| T_0Y_0\right| }\left| (c_t\pm
c_x)/c_y\right| \quad\exp (-(c_t-c_y)\tau /2) \\
&  &  \\
A_3 & \rightarrow & i\sqrt{\left| T_0/Y_0\right| }\left| (c_t\pm
c_x)/c_y\right| \quad\exp (-(c_t+c_y)\tau /2).
\end{array}
\end{equation}
We can see that these limits correspond to Bianchi I solutions, that is,
Bianchi II solutions approximate (different) Bianchi I solutions as $\tau
\rightarrow \pm \infty $. These Bianchi I solutions differ in the value
of the discrete parameter $\epsilon $ and renormalizations of $
T_0,Y_0$.

As in the case of Bianchi type I, the pull-back of the symplectic form to
the space of real solutions is:

\begin{equation}  \label{II:realsymp}
\hat \Omega =dc_t\wedge d\ln T_0+dc_y\wedge d\ln Y_0,
\end{equation}
where $c_t,c_y,T_0,Y_0$ are Dirac observables for the theory.

\section{Characterization of the ``bounce''}

When solving Einstein's equations in the Hamiltonian formulation for the
``diagonal models'', the usual geometrodynamical description \cite{ryanshep}
is in terms of variables $\{\beta ^A,p_A\},$ with the index $A$ taking
values $0,\pm 1$ where $\beta ^0$ is the log of the 3-volume and $\beta
^{\pm }$ are the anisotropies. The Hamiltonian constraint has a term
quadratic in momenta ($\eta ^{AB}p_Ap_B$) and a potential term which for all
Bianchi types (except Bianchi I) has exponential dependence on the
2-dimensional $\beta ^{\pm }$ plane. Therefore the dynamics is that of a
particle (the ``universe point'') moving on a flat background in the given
potential. In the regions where the potential vanishes the particle moves
essentially free along a (null) straight line as if it were a solution to
Bianchi I equations (Kasner epoch), until it encounters a potential wall and
bounces back along a (reflected) straight line. In Bianchi II there is only
one ``wall'' in the potential, and therefore the world-particle bounces once
between the initial singularity and the expanding evolution. In Bianchi IX
models, there are 3 walls and in general the particle keeps bouncing within
the walls,{\ approaching the singularity as $t\rightarrow 0$. }

This classical behaviour was first analyzed by Belinskii et al\cite
{belinskii}. They use the Bianchi II solution as a transition between Kasner
epochs in the general Bianchi IX solution. In each Kasner epoch, the
solution is approximately the Kasner solution, where the metric component in
one spatial direction decreases with time (it has a negative Kasner
exponent) and the other two increase with time (they have positive
exponents). In the transition between Kasner epochs, the negative power of
time is transferred from one spatial direction to another, so one metric
component reaches a minimum and another a maximum, while the third one
increases monotonically during the transition. Bianchi II models have a
single transition, and the ``bounce'' is then defined as the point along the
trajectory when one of the metric components reaches its maximum (which is
different from the point when the other metric component reaches its
minimum).

Now, if we consider the Hamiltonian formulation in terms of the new
variables, the Hamiltonian constraint contains just a term quadratic in
momenta. Therefore, the dynamics can be interpreted as that of a free
particle moving on a null geodesic in the complex 3-dimensional space $%
\{A_I\}$, where the supermetric is given by $G^{IJ}$ in (\ref{gensc}). Since
there is no potential, there is neither a wall nor any particular coordinate
time to identify with the ``bounce'' time that characterizes Bianchi II in
the geometrodynamical formulation. If we translate the bounce time defined
in the geometrodynamical variables, we find that it is the point on the
trajectory where the function $t(\tau )$ vanishes, that is, when the Killing
vector $K_t $ vanishes. However, since this description is in terms of
Killing symmetries that do not exist in some of the higher Bianchi models,
this characterization of ``bounce'' times is not useful for the study of the
other models.

Since the dynamics is determined completely by the supermetric $G^{IJ}$, one
could look for an invariant description of the bounce in terms of the {\em %
geometry} of the curved superspace in which the ``universe point'' moves.
Any scalar quantity that we construct from the supermetric is a function
only of the coordinate $x(\tau )$, which when evaluated along real
trajectories depends only on the function $\cosh (c_x(\tau -\tau _0))$.
Therefore there is a symmetry that one can identify: $(\tau -\tau
_0)\rightarrow -(\tau -\tau _0)$. Again, however, this symmetry does not
survive when we consider other Bianchi models. Hence, we would like to find
a way of describing a turn-around point that does not depend on the use of
symmetries particular to the Bianchi II solutions. One possibility is to
look for features such as maxima or singularities in (the absolute values
of) various superspace curvature scalars like the Ricci scalar $R_{IJ}G^{IJ}$
, or $R_{IJ}R^{IJ}$. Unfortunately, all the real metric scalars up to second
order in the Ricci tensor have several extrema as functions of $\tau $ when
evaluated on the real trajectories, and therefore do not uniquely identify
the bounces in higher Bianchi models, like type IX.

There is however, another option. The Bianchi II solutions behave
asymptotically ($\tau \rightarrow \pm \infty $) like specific Bianchi I
solutions. On real Lorentzian Bianchi I solutions, the connection components
are pure imaginary, exponential functions of $\tau $. Hence we can look for
a characterization of the ``transition region'', by studying the ratio of
real to imaginary parts of the solutions for the connection components in
Bianchi type II: we know that this ratio tends to zero in the asymptotic
regions as $\tau \rightarrow \pm \infty $, and since the Bianchi II
connections are in general complex, this ratio is non-zero at finite $\tau $.
The functions $\Phi _I=Re(A_I)/Im(A_I)$ turn out to be very simple:
\begin{equation}  \label{phaseA}
\begin{array}{rcl}
\Phi _1^{-1} & =&\epsilon \delta \sinh (c_x(\tau -\tau _0)-\phi ) \\
& & \\
\Phi _2^{-1} & =\epsilon \cosh (c_x(\tau -\tau _0)-\phi /2)/\sinh (\phi /2)
\quad {\rm if }\quad \delta =+1 \\
& =-\epsilon \sinh (c_x(\tau -\tau _0)-\phi /2)/\cosh (\phi /2) \quad {\rm
if }\quad \delta =-1 \\
&  \\
\Phi _3^{-1} & =\epsilon \sinh (c_x(\tau -\tau _0)-\phi /2)/\cosh (\phi /2)
\quad {\rm if }\quad \delta =+1 \\
& =-\epsilon \cosh (c_x(\tau -\tau _0)-\phi /2)/\sinh (\phi /2) \quad {\rm
\quad if }\delta =-1,
\end{array}
\end{equation}
where we recall that $\delta =sgn(c_tc_y)=-sgn(T_0Y_0)$ and $\phi =
\mbox{arctanh} (c_x/c_t)$. Each of these functions has a unique and
very obvious ``transition point'' (where they either have a maximum or
diverge) at either $c_x(\tau -\tau _0)=\phi $ (for $\Phi _1$) or
$c_x(\tau -\tau _0)=\phi /2$ (for $\Phi_2,\Phi_3$). The zero of $\Phi
_1^{-1}$ is exactly the ``bounce time'' as defined in
\cite{belinskii}. Thus, the ``Kasner epochs'' in the Ashtekar
variables are identified with pure imaginary, exponential connections
and the ``bounces''\ are either maxima or divergences of the (tangent
of the) phase of the connection components, or in other words, the
most extreme departure from Bianchi I connections. Both before and
after the bounce, the connections have an asymptotic exponential
behavior, but with different logarithmic velocities. This description
is very similar to that by Belinskii {\it et al} in terms of the
behaviour of metric components. As in the geometrodynamical
description where the maximum of one metric component does not
coincide with the minimum of the other (forbidding an invariant
definition of the bounce time), here too the divergences and maxima in
the phase of the connection components do not coincide.

\section{Conclusion}

Let us briefly recapitulate the results we have obtained. For both Bianchi
types I and II, we first solved the {\em complex} equations of motion by
finding the null geodesics of the complex supermetric on the space of
connections. The space of solutions is parametrized by 4 complex variables,
which form 2 canonically conjugate pairs. Next, to find the real Lorentzian
solutions, we required that the triads are real throughout the real
evolution. This condition is satisfied when the parameters on the space of
solutions are all real and satisfy in addition certain nonholonomic
constraints. (Note that we have made a simplifying choice of variables in
order to obtain a real canonical parametrization of the reduced phase space.)

All Bianchi II solutions approach a (different) Bianchi I solution
asymptotically in the past or the future, and deviations from Bianchi I like
behaviour and the transition from one asymptotic Bianchi solution to another
occurs at some finite time. Now for Lorentzian Bianchi I, $Re(A_I(t))=0$,
whereas the Lorentzian connections for Bianchi type II are in general
complex. So it is particularly nice to characterize the bounce in Bianchi II
as a deviation from Bianchi I like behaviour via maxima or divergences in $%
\Phi _I=Re(A_I(t))/Im(A_I(t))$. We have seen that for any given solution,
one of the phases $\Phi _I$ has a maximum while the other two have a
singularity. This does uniquely characterize the bounce and could be used
for the counting of bounces in the numerical study of Bianchi IX in the new
variables.

Recall that we failed to find a description independent of the coordinates
on the (minisuper)space of connections. In retrospect it is not surprising
that there is no characterization in terms of some supermetric curvature
scalars, since the Lorentzian solutions have not been described as null
geodesics in some real supergeometry, though the complex solutions are null
geodesics in a complex superspace.

One can proceed to construct the reduced space quantum theory for these
models. This construction is straightforward since the reduced phase space
is coordinatized by two pairs of real canonically conjugate variables.
However, from the point of view of learning something about the Dirac
quantum theory in terms of the connections themselves, this is not a very
useful approach. In particular, one does not see clearly the role of the
complex reality conditions in finding an inner product on the space of
solutions to the quantum scalar constraint equation\cite{aa:rt}, or whether
this is even feasible. These issues will be explored in a later paper\cite
{gmm:rst}, in which we construct the complete Dirac quantum theory for the
Bianchi II model. The classical Dirac observables $(c_t,c_y,T_0,Y_0)$ we
have constructed here do have quantum analogs, and the Hermiticity
conditions on them serve to select an inner product uniquely on the space of
solutions to the quantum scalar constraint equation.

\section{Appendix: Special case solutions}

For special values of the parameters $c_y, c_t$, the limits of the general
solutions we found in section 4 are not themselves solutions to the limiting
equations of motion. In order to find the correct solutions at these
limiting values, we have to solve directly the reduced equations of motion
obtained by taking the appropriate limits of the general ones. There are
three special cases and we consider them one by one.

\subsection*{Case 1: $c_t=0$}

In this limit, the equations of motion (\ref{II:ci},\ref{cxeqn}) reduce to
\be
\begin{array}{rl}
\dot t & =0 \\
{\dot y}/y & =c_y(1+ 1/x) \\
\dot x^2 & =c_y^2(1-x^2).
\end{array}
\ee
The solution to the above equations is
\be
\begin{array}{rl}
t(\lambda ) & =i\epsilon t_0 \\
y(\lambda ) & =iy_0e^{c_y\lambda } {\displaystyle \frac{\sin (c_y(\lambda
-\lambda _0))}{1+\epsilon \cos (c_y(\lambda -\lambda _0))}} \\
x(\lambda ) & =\epsilon \sin (c_y(\lambda -\lambda _0)).
\end{array}
\ee
Note that the order and the form of the limiting equations are the same as
in the general case, and thus we expect that the above solution to the
limiting equations can be obtained by taking the limit ($c_t\mapsto 0$) of
the general solution (\ref{II:xsol},\ref{II:orbits}), as can be confirmed by
direct calculation. Thus we can directly take the limits of the reality
conditions\ (\ref{II:realconds}), and conclude that $c_y=0$. The triads $E_I$%
's in this case (with $c_t=c_y=0$) vanish, and the connections $A_I$'s are
constants.

\subsection*{Case 2: $c_y=0$}

The equations of motion in this case are:
\be
\begin{array}{rl}
{\dot t}/t & =c_tx/ {(1-x)} \\
\dot y & =0 \\
\dot x^2 & =c_t^2x^2.
\end{array}
\ee
and the solution is
\be
\begin{array}{rl}
x(\lambda ) & =e^{\epsilon c_t(\lambda -\lambda _0)} \\
y(\lambda ) & =y_0 \\
t(\lambda ) & =t_0(x(\lambda )-1)^{-\epsilon }.
\end{array}
\ee

If we take $\epsilon =-1$, then $E_2=E_3=0$ and $E_1=ic_t/(G_Nt_0)$, and the
connections are $A_1=t_0,\; A_2=y_0,\; A_3=\sqrt{t_0y_0}\exp (-c_t(\lambda
-\lambda _0)/2)$. The reality conditions only require that $c_t/t_0$ is pure
imaginary.

If we take $\epsilon =+1$, then the triads are:
\be
\begin{array}{rcl}
E_1 & = & {\displaystyle \frac{ic_t}{G_Nt_0}}(1-e^{2c_t(\lambda -\lambda
_0)}) \\
E_2=E_3/y_0 & = & {\displaystyle \frac{2ic_t}{\sqrt{t_0y_0}}}e^{c_t(\lambda
-\lambda _0)/2}.
\end{array}
\ee

The reality conditions then require $c_t,it_0,y_0\in \real
$ , $\cos (c_t(\phi -\phi _0))=0$, and $it_0y_0\sin (c_t(\phi -\phi _0))>0$.
Let us define $\delta =\sin (c_t(\phi -\phi _0))=\pm 1,$ and $T_0=i\delta
t_0,Y_0=y_0$ (so $T_0Y_0>0$). The Lorentzian solutions for triads and
connections are, then:
\be
\begin{array}{rcl}
E_1 & = & - {\displaystyle \frac{\delta c_t}{T_0}}(1-e^{2c_t(\tau -\tau _0)})
\\
E_2=E_3/Y_0 & = & {\displaystyle \frac{2c_t}{\sqrt{T_0Y_0}}}e^{c_t(\tau
-\tau _0)/2} \\
&  &  \\
A_1 & = & i\delta T_0(1-i\delta \exp (c_t(\tau -\tau _0)))^{-2} \\
A_2=Y_0A_3 & = & i\sqrt{T_0Y_0}{\displaystyle \frac{\exp (c_t(\tau -\tau _0))%
}{1-i\delta \exp (c_t(\tau -\tau _0))}}.
\end{array}
\ee
Notice that although the triads have the same form as (\ref{II:realtriads})
(ignoring constant multiplicative factors, and taking $c_y=0,c_x=c_t$ in the
exponentials), the connections do not have the same behavior as in (\ref
{II:realconn}). In other words, the limit $c_y\rightarrow 0$ can be taken
smoothly in the metric variables, but not so in the connection space.

\subsection*{Case 3: $|c_y|=|c_t|$}

We set $c_y=\delta c_t$, where $\delta =\pm 1$. The equations of motion are:
\be
\begin{array}{rl}
{\dot t}/t & =c_tx/( {1-x}) \\
{\dot y}/y & =\delta c_t(1+1/x) \\
\dot x{}^2 & =c_t^2,
\end{array}
\ee
and yield the solution
\be
\begin{array}{rl}
x(\lambda ) & =\epsilon c_t(\lambda -\lambda _0) \\
y(\lambda ) & =y_0(\lambda -\lambda _0)^{\delta \epsilon }e^{\delta
c_t\lambda } \\
t(\lambda ) & =t_0(1-\epsilon c_t(\lambda -\lambda _0))^{-\epsilon
}e^{-c_t\lambda }.
\end{array}
\ee
We see that $\delta \rightarrow -\delta $ is equivalent to $y\rightarrow
y_0^2/y$, and this in turn is equivalent to $\{A_2\longleftrightarrow
y_0^{}A_3,\;E_2\longleftrightarrow y_0E_3\}$, so we only need to consider in
detail the case $\delta =1$.

If $\epsilon =+1$, the triads are:
\be
\begin{array}{rcl}
E_1 & = & {\displaystyle \frac{-2ic_t}{t_0}}e^{c_t\lambda }(1-c_t(\lambda
-\lambda _0)) \\
E_2 & = & \sqrt{{\displaystyle \frac{c_t}{t_0y_0}} }{\displaystyle \frac{%
1+c_t(\lambda -\lambda _0)}{ \lambda -\lambda _0}} \\
E_3 & = & \sqrt{{\displaystyle \frac{c_ty_0}{t_0}}}e^{c_t\lambda
}\;(1+c_t(\lambda -\lambda _0)).
\end{array}
\ee
We can see by inspection that the only solution to the reality conditions is
the trivial one, $c_t=0$, (vanishing triads and constant connections).

If $\epsilon =-1$, then the triads are:
\be
\begin{array}{rcl}
E_1 & = & 0 \\
E_2 & = & i \sqrt{{\displaystyle \frac{c_t}{t_0y_0}}}\sqrt{{\displaystyle
\frac{1+c_t(\lambda -\lambda _0)}{ 1-c_t(\lambda -\lambda _0)}}} \\
E_3 & = & i\sqrt{{\displaystyle \frac{c_ty_0}{t_0}}}{\displaystyle \frac{
e^{c_t\lambda }}{\lambda -\lambda _0}}\sqrt{{\displaystyle \frac{%
1+c_t(\lambda -\lambda _0) }{1-c_t(\lambda -\lambda _0)}}}.
\end{array}
\ee
Here, the reality conditions are satisfied when $\phi =\phi _0=0$ and $%
c_t,t_0,$ and $y_0$ are real and satisfy $c_tt_0y_0<0$.

{\bf Acknowledgments:} The authors would like to thank Abhay Ashtekar for
suggesting the research topic and for many useful discussions, Paul Tod
and Jorge Pullin for discussions at various stages
and the members of the Center for Gravitational
Physics and Geometry for their hospitality. This work was supported in part
by NSF grants PHY 93-96246, PHY 90-08502, PHY 90-16733 and PHY 91-13902,
funds provided by Syracuse University and the University of California,
Santa Barbara and by the Eberly research fund of Pennsylvania State
University.

\end{document}